\documentclass{aastex}
\usepackage[dvips]{epsfig}
\begin{document}

\title{Hubble Space Telescope Observations of the Gravitationally
Lensed Cloverleaf Broad Absorption Line QSO H1413+1143: Imaging
Polarimetry and Evidence for Microlensing of a Scattering Region$^1$}

\author{Kyu-Hyun~Chae,\altaffilmark{2,3,4}
David~A.~Turnshek,\altaffilmark{2,3}
Regina~E.~Schulte-Ladbeck,\altaffilmark{2,3}
Sandhya~M.~Rao,\altaffilmark{2,3} and
Olivia~L.~Lupie\altaffilmark{2,5}}

\altaffiltext{1}{Based on observations made with the NASA/ESA
$Hubble~Space~Telescope$, obtained from the Space Telescope
Science Institute, which is operated under NASA contract NAS 5-26555.}
\altaffiltext{2}{email: chae@jb.man.ac.uk, turnshek@pitt.edu, rsl@pitt.edu,
rao@phyast.pitt.edu, lupie@stsci.edu}
\altaffiltext{3}{Dept. of Physics \& Astronomy,
University of Pittsburgh, Pittsburgh, PA 15260, USA}
\altaffiltext{4}{Present address: University of Manchester, JBO,
Macclesfield, Cheshire SK11 9DL, UK}
\altaffiltext{5}{Space Telescope Science Institute, 3700 San
Martin Drive, Baltimore, MD 21218, USA}

\begin{abstract}
We report the results of {\it Hubble Space Telescope} Wide Field and
Planetary Camera 2 broadband F555W and F702W photometric and F555W
polarimetric observations of the ``Cloverleaf'' QSO H1413+1143. This
is a four-component gravitationally-lensed broad absorption line
(BAL) QSO. Observations were obtained at two epochs in March 1999
and June 1999 separated by $\approx 100$ days. The observations were
photometrically and polarimetrically calibrated using the standard
``pipeline'' calibration procedures implemented at the Space
Telescope Science Institute. The goal of our program was to detect
any {\it relative} changes among the components and between the
two epochs.  Over this time baseline we detected an $\approx 0.07$
mag dimming in component D of the lensed image, which we interpret
as evidence for microlensing.  In March 1999 we find significant
evidence for a difference in the relative linear polarization of
component D in comparison to the other three components; in June
1999 the combined polarization of the Cloverleaf components was
lower.  In March 1999 the apparently microlensed component D has a
rotated polarization position angle and a somewhat higher degree of
polarization than the other three components.  We suggest that this
difference in polarization is due to microlensing magnification
of part of a scattered-light (i.e. polarized) continuum-producing
region. The results indicate that in the Cloverleaf the size-scale
of the polarized scattered-light region exceeds $\approx 10^{16}$
cm but lies interior to the region producing the broad emission lines
($< 10^{18}$ cm).

\end{abstract}


\section{Introduction}

Broad absorption line (BAL) QSOs comprise $\approx$ 10\% of the objects
in optically selected QSO samples. Their defining characteristics are
deep, high-velocity (usually $< 0.1c$) absorption troughs blueward
of high-ionization broad emission lines (BELs) in species such as
\ion{Si}{4}, \ion{C}{4}, \ion{N}{5}, and \ion{O}{6} (e.g.\ Turnshek
1988 and Weymann et al.\ 1991).  They also have a distribution of
linear polarizations which peaks at a significantly higher polarization
in comparison to non-BAL radio-quiet QSOs (Schmidt \& Hines 1999;
Hutsem\'{e}kers, Lamy \& Remy 1998; Goodrich 1997; Turnshek 1988);
indeed the origin of much of the polarization in most non-BAL QSOs is
probably the Galactic interstellar medium.

The observed polarization properties of BAL QSOs depend on, and
therefore contain relevant astrophysical information on, the geometries
and physical properties of the inner regions of these QSOs (i.e. the
narrow emission-line region, the dusty torus, the BAL region, the BEL
region, any scattering regions, the thermal accretion disk, and any
region producing non-thermal emission).  Recent studies 
on the polarization properties of BAL QSOs (Schmidt \& Hines 1999; 
Hutsem\'{e}kers et al. 1998; Ogle 1998; Goodrich 1997) have 
shown that:
(1) their continuum polarizations are, on average, significantly higher
than non-BAL radio-quiet QSOs, with the degree of polarization
rising mildly toward shorter wavelengths, (2) the BALs are more highly
polarized than the continuum, with position angle rotations observed in
the BAL troughs, (3) BELs in {\it some} BAL QSO spectra are polarized,
but the degree of polarization is lower than in the continuum and the
polarization position angles are not necessarily similar, and (4) there
is some evidence that the degree of polarization is positively correlated
with the BAL QSO's balnicity index (defined by Weymann et al.\ 1991)
and the presence of low-ionization BALs, with objects having 
higher balnicity indices and/or low-ionization BALs being 
more polarized on average.

While it is generally agreed that the scattered continuum is, in large
part (if not totally), responsible for the observed net polarizations
observed in BAL QSOs (e.g.\ Goodrich \& Miller 1995; Ogle 1998; Schmidt
\& Hines 1999), the size-scales and geometries of the regions containing
scattering particles (i.e.\ electrons and/or dust particles) are not yet
well constrained. However, the gravitationally-lensed
Cloverleaf BAL QSO H1413+1143 ($z_{em} \approx 2.55$) is at present a
unique laboratory for study of the polarization mechanism. Its 
four components are known to have a combined net continuum polarization
which has varied between $1.5 - 3.5$\% over a decade (Goodrich \& Miller
1995) and one of the four components (component D) shows evidence for
microlensing in the form of light variability (see Angonin et al.\ 1990, 
Arnould et al.\ 1993, Remy et al.\ 1996, Ostensen et al.\ 1997, 
and new evidence presented here) and differences in BEL and BAL profile 
characteristics (Chae \& Turnshek 1999 and references therein).

We know that, in principle, microlensing could alter the net
polarization of a single gravitationally-lensed component, since
this process selectively produces additional magnification of a small
region of an Einstein ring radius on the source plane. The resulting
polarization properties during microlensing would depend on the detailed
geometry of the region where the polarized light originates (see 
Belle \& Lewis 2000 for
examples). Consequently, the observed component polarization properties
during microlensing can be used to constrain the projected sizes of the
scattering regions and their projected distances from any central black
hole (i.e.\ the central region of the accretion disk).

In this {\it Letter}, we report the results of {\it Hubble Space Telescope
(HST)} Wide Field and Planetary Camera 2 (WFPC2) broadband F555W and
F702W photometric and F555W polarimetric observations of the Cloverleaf.
We find new evidence for microlensing of component D and show that
component D has significantly different relative polarization properties than
the other three components.  In \S2 we describe our observations and
data analysis; in \S3 we present the results; and in \S4 we discuss the
implications for models of the production of polarized light in BAL QSOs.
We note that this paper is the fifth in a series of {\it HST} results
on the Cloverleaf by members of our group.  Earlier results include
constraints on the sizes and shapes of absorbing regions producing the
BALs (Turnshek 1995), constraints on the component image locations and
magnifications (Turnshek et al. 1997), constraints on the properties
of the intervening absorbers seen in the component spectra (Monier,
Turnshek \& Lupie 1998), and implications for gravitational-lens models
of the Cloverleaf, including a constraint on BEL region size scales
(Chae \& Turnshek 1999).

\section{Observations and Data Analysis}

For this study the Cloverleaf QSO was observed at two closely-spaced
epochs, 15-16 March 1999 and 23-24 June 1999. Broadband F555W filter
observations were made with the Wide Field Camera 2 (WFC2) with either 
no polarizer or one of four polarizers (POLQ, POLQN18, POLQN33, POLQP15).
Also, F702W filter observations were made with the Planetary Camera 2
(PC2) and no polarizer.  The observations are summarized in Table~1
and numbered as a referencing convenience. Each
F555W observation consisted of taking two sets of four dithered images.
The pixel position of component A in the second set was chosen to coincide
with that of component D in the first set. In addition to the normal
procedures for identifying cosmic ray contamination of WFPC2 images, all
pipeline processed images were also individually examined. Cosmic
rays present near the Cloverleaf components were removed interactively
and the data were replaced using interpolation.  After sky subtraction,
each set of images was combined via the variable-pixel
linear reconstruction algorithm, or the ``drizzling''
algorithm (Fruchter \& Hook 1998; Fruchter et al.\ 1997). Point
spread function (PSF)-fitting photometry showed that the photometric
results of the two sets for each observation were consistent with each
other. Finally, all eight images in the two sets were drizzled to form
one image. The procedure of taking dithered images and combining them
via drizzling partially restores the PSF, thereby facilitating more
accurate PSF-fitting photometry.  Using the final drizzled image for
each observation, an empirical PSF was constructed iteratively from the
Cloverleaf components themselves, and then the Cloverleaf components were
fitted simultaneously using this empirical PSF.\footnote{We 
could not use the ``Tiny TIM'' software to generate PSFs since 
it does not have the capability of incorporating the effects of the polarizing
filters. We found that all individual images taken with
the POLQN33 polarizer (for both epochs) and the POLQP15 polarizer (for
the March epoch) had substantially modified PSFs in terms of FWHM and
peak pixel value in comparison to images taken without polarizers.} For
the above procedures, the Image Reduction and Analysis Facility (IRAF)
packages DAOPHOT and STSDAS were used. The reliability of the above
PSF-fitting photometry was tested by visually examining the PSF-subtracted
region and computing pixel statistics within the subtracted region; the mean
pixel data number ({\it DN}) value within the region was $|\overline{DN}|
\lesssim 0.1\sigma$ in all cases.  

The observations were photometrically and polarimetrically
calibrated using the standard ``pipeline'' calibration procedures
implemented at the Space Telescope Science Institute. The goal of
our program was not to make absolute measurements, but to detect any
relative changes among the components and between the two epochs.
Thus, the errors we quote based on our differencing procedures are
statistical in nature. Our relative measurements and the resulting
interpretation should not be affected by a small systematic error 
in the pipeline calibration.

The derived F555W photometric results without polarizers (i.e.\
observation numbers 1, 2, 3 and 11, 12, 13) 
were found to be highly consistent with one another.  This served
as an independent test, confirming the reliability of our method of photometry.
The F702W photometric observations without polarizers (i.e. \
observation numbers 4, 5, 6 and 16, 17, 18) 
allowed us to fill up the remaining time
available for exposures in a number of the orbits
using a filter that had previously been used to observe the Cloverleaf,
but without the goal
of deriving photometry using drizzling.  We found the statistical
errors to be considerably reduced using the drizzling method.

In order to determine the Stokes parameters (I, Q, U), the F555W
observations with polarizers were incorporated into the WFPC2 polarization
calibration model (Biretta \& McMaster 1997). The Stokes parameters
are related to the degree of polarization ($p$) and the position
angle ($PA$) via the relations $p = (u^2 + q^2)^{1/2}$ and $PA = (1/2)
\tan^{-1}(u/q) + n\pi/2$ where $q \equiv Q/I$, $u \equiv U/I$, and $n =
0,1,2$ for $u \geq 0$ and $q \geq 0$, $u \geq 0$ and $q \leq 0$ (or, $u <
0$ and $q < 0$), and $u \leq 0$ and $q \geq 0$, respectively.  For the
March epoch, six observations
were incorporated simultaneously using a $\chi^2$ fitting technique to
determine the Stokes parameters.\footnote{For the March data we did
not use the web-based WFPC2
polarization calibration tool or the IRAF $IMPOL$ package, since the
former can take only three relative counts at one time while the latter
takes the images themselves as input using only aperture photometry,
which is less accurate for the ``crowded'' Cloverleaf field. It was
possible to use the web-based WFPC2 polarization calibration tool for
the June data.} The $\chi^2$ is defined by
\begin{equation}
\chi^2 = \sum_i \left( \frac{C_i^{\mbox{\scriptsize obs}} - 
             C_i^{\mbox{\scriptsize mod}}(I,Q,U)}{\sigma_i} \right)^2,
\end{equation}
where $i = 7,8,9,10,11,12$ are the March observation numbers, 
$C_i^{\mbox{\scriptsize obs}}$  and $C_i^{\mbox{\scriptsize
mod}}$ are the observed and calibration-model-predicted
counts, respectively, and $\sigma_i$ are the statistical
errors in the observed counts.  The minimum $\chi^2$ value was
$\chi_{\mbox{\scriptsize min}}^2 < 7$, with 3 degrees of freedom
for all components, indicating that there is a reasonable match
between the model and the data. The $1\sigma$ statistical errors
for each Stokes parameter were estimated using $\Delta\chi^2
=\chi^2 -\chi_{\mbox{\scriptsize min}}^2 = 1$ (e.g.\ Press et al.\
1992), and the statistical errors for $p$ and $PA$ were estimated
assuming Gaussian error propagation.  For the June epoch ($i = 19, 20, 21$), 
the Stokes parameters were determined by solving $C_i^{\mbox{\scriptsize obs}}
= C_i^{\mbox{\scriptsize mod}}(I,Q,U)$ for I, Q
and U,$^5$ with the statistical errors for the Stokes parameters 
being estimated using Gaussian error propagation of the photometric
errors via these equations.

The results for the June epoch are much less reliable in comparison
to the March results.  There are two reasons for this. First,
the June measurements were made at only three polarization angles,
while the March measurements were made at six polarization angles.
Second, according to the pipeline calibration, the degree of
polarization is evidently smaller in June 1999, which gives rise to
a lower signal-to-noise ratio for a fixed photometric error. Thus,
the June results should not be given high weight. Since the March
data show the polarization of components A, B and C to be similar,
for the June epoch we report only the results for the combined 
measured polarization of components A, B and C, and we report the
measured polarization of component D separately.

\section{Results}

Below we consider the results of our WFPC2 observations in two
separate parts. First, we consider any evidence for photometric
variations among the four lensed components of the Cloverleaf
between the June 1999 and March 1999 epochs (\S3.1, Table 2a,b).
Second, we consider any evidence for differences in the linear
polarization among the four lensed components (\S3.2, Table 3).

\subsection{Brightness Variation of Component D: Evidence for Microlensing}

The photometric observations of the Cloverleaf without polarizers
were used to search for and measure any brightness variations over the
time baseline ($\approx 100$ days) which separated the two epochs of
observation.  As noted earlier, the individual F555W photometric results
for each epoch were very consistent with one another; the average F555W
photometric results for each epoch are reported in Table 2a,b. The average
F702W photometric observations are also reported in Table 2a,b, but are less
accurate. The photometric results are shown in Figure 1. They clearly
indicate that component D decreased its brightness over the $\approx 100$
day time baseline, while the other three components either remained about
the same brightness (i.e.\ components B and C) or slightly increased their
brightness (i.e.\ component A).  Assuming an insignificant wavelength
dependence to the brightness variation over the F555W and F702W passbands,
using variance-weighting we find $\Delta m_D = 0.074\pm 0.004$ mag and
$\Delta m_A = -0.023\pm 0.004$ mag. Evidence for brightness variations in
components B and C are at the 2$\sigma$ level of significance or less.

Recent detailed work on modeling the Cloverleaf lens and time delays
(Chae \& Turnshek 1999) suggests that in all reasonable models component
C is the leading component and component D is the trailing component,
with the maximum time delay being model-dependent and lying in the
range $\approx 7-41$ days.  However, the
predicted time delays between components C and A or components C and
B are always a significant fraction (at least $\approx 30$\%) of the
predicted time delay between components C and D.  Given these lens models
and observations of brightness variations in other BAL QSOs (Sirola et
al. 1997), we believe that these new photometric data on the Cloverleaf
are not likely to be consistent with simply an intrinsic variation in
the source BAL QSO's brightness coupled with time delays among the four
components.  The most likely cause of the light variation of component
D is microlensing, which is consistent with the findings of others who
have studied brightness variations in the Cloverleaf (Angonin et al.\
1990; Arnould et al.\ 1993; Remy et al.\ 1996; Ostensen et al.\ 1997).

\subsection{Polarimetric Results}

For the March 1999 and June 1999 epochs of observation, Table
3 gives the measured normalized Stokes parameters $q$ and $u$,
the degree of linear polarization ($p$), the corrected degree of
linear polarization ($p_{\mbox{\scriptsize corr}}$) which takes
into account the bias toward measuring higher polarization in low
signal-to-noise data (Wardle \& Kronberg 1974), and the polarization
position angle ($PA$).  For the March observations the derived $q$
and $u$ Stokes parameters are shown in Figure 2, and it is seen
that the polarizations of components A, B and C differ from the
polarization of component D.  

The main points of the polarimetric results are: (1) The
polarizations of components A, B, and C in March 1999 have no
appreciable differences.  (2) As is clearly seen in Figure~2,
in March 1999 component D has a significantly different relative 
polarization in comparison to the other three components. The 
{\it relative} normalized Stokes parameters between component 
D and the combination of the other components in March 1999 is 
$\Delta q_{\mbox{\scriptsize D,$\overline{\mbox{ABC}}$}} 
= -1.26\pm 0.54\%$ and 
$\Delta u_{\mbox{\scriptsize D,$\overline{\mbox{ABC}}$}} 
= -2.32\pm 0.61\%$ (Table~3), which is a difference in polarization 
at a level of significance of 4.5$\sigma$. (3) The pipeline calibrated 
data suggest that the net polarization of the Cloverleaf changed 
between the March 1999 and June 1999 epochs, with the polarization being
smaller during the June epoch.

\section{Discussion} 

These results are the first observational
ones which address resolved polarization measurements in a
gravitationally-lensed QSO and, owing to our interpretation (\S3.1
and below), the first to report evidence for microlensing of
a polarized-light region in a QSO.

Before we examine the implications of our interpretation, we
should comment on some issues which can affect the interpretation.
For example, polarization induced by any dust which is present along
the sight-line toward the Cloverleaf is unlikely to be responsible
for the added component of polarization that appears to be present
in component D.\footnote{Interstellar polarization within the Milky
Way is not significant for the Cloverleaf (e.g.\ Hutsem\'{e}kers et
al.\ 1998).} There are several reasons for this. First, multicolor
(F336W, F702W, F814W) {\it HST} WFPC2 observations have shown
that differential dust reddening is, in fact, present across
the Cloverleaf components.  Component B is the most reddened and
component C is the least reddened, with the F336W$-$F814W color index
being $0.56\pm 0.04$ mag between components B and C.  The reddening
of components A and D are intermediate (see figure 3 in Turnshek et
al.\ 1997). The source of this differential reddening may be dust
in the interstellar medium of the lensing galaxy. However, since
interstellar polarization is normally proportional to the amount
of reddening, we would not expect the change in polarization of
component D to be related to the reddening.  Second, if the source
of the polarization was the lens, it would seem unlikely that the
induced polarization would be different only for component D. Third,
interstellar polarization would not be expected to be time-variable,
but the observations indicate that the polarization changed between
the March 1999 and June 1999 epochs.  Fourth, Faraday rotation
due to Galactic or cosmologically intervening plasma that may be
present along the sight-line to component D could not reasonably give 
rise to any rotation in the polarization position angle.  The observed
change in position angle for component D, $\Delta \theta = 32 \pm 7$
deg at $\lambda = 5300$ \AA\ (for the F555W filter), would require
a medium with a rotation measure of ${\mbox{RM}} = (2.0 \pm 0.4)
\times 10^{12}$ radians~m$^{-2}$; however, for example, this is
eight orders of magnitude larger than one of the highest values ever
measured for an extragalactic radio source (e.g. 3C 295, Perley \& 
Taylor 1991).

Consequently, we have argued that the $\approx 0.07$ mag relative
decrease in brightness of component D over the $\approx 100$ day
interval between March 1999 and June 1999 (Figure 1) is evidence for
microlensing of component D.  In this scenario, component D evidently
faded due to the motion of a microlens, causing it to be
less magnified in June 1999.  This is consistent with the
earlier conclusions of Chae \& Turnshek (1999, also \S3.2), who
interpreted the lower equivalent widths of the BELs in component
D (observed with HST FOS in both June 1993 and December 1994)
as evidence for microlens magnification of just the continuum
of component D (not the BELs).  Now, relying on the new results
presented here, we can refine some of the conclusions of Chae \&
Turnshek (1999), and place better qualitative constraints on the
size-scale of the polarized scattered-light region in relation to our
understanding of the size-scales of the BEL and continuum-producing
regions in QSOs.

We should point out that in all published individual spectra of
the Cloverleaf components, the equivalent widths of the BELs in
component D are smaller than observed in the other components. The
first of four sets of component spectra were obtained in the spring
of 1989 (Angonin et al. 1990) and this trend continues up until at
least the spring of 2000, when HST-STIS spectra taken by Monier
(PI) and several of the authors continued to show component D to
have BELs with lower equivalent widths. This suggests that some
level of microlens magnification of the continuum-producing region
(but not the BELs) of component D is common over a relatively long
time baseline.

The Cloverleaf's macrolens by itself achromatically amplifies all
light lying within $\approx 10^{19-20}$ cm of the central source
into four point-like image components (see figure 1 of Chae \&
Turnshek 1999).  Observations indicate that the macrolensed region
includes the BEL region. This is consistent with expectations since
the size of the region producing BELs like \ion{C}{4} and Ly$\alpha$
is estimated to be $\approx 10^{18} L_{46}^{0.5}$ cm from the central
photoionizing source (Murray \& Chaing 1998; Kaspi et al. 2000),
where $L_{46}$ is the lensed QSO luminosity in units of $10^{46}$
ergs s$^{-1}$. The source QSO luminosity in the Cloverleaf is not
well-constrained because observations only provide results on the
relative component amplifications; however, the lens models suggest
that the luminosity is likely to be of order $L_{46}$.  Evidently
the Cloverleaf is polarized because an asymmetric
continuum scattering region also lies within the macrolensed region.
The asymmetry is a requirement since the net polarization is non-zero. 
This region is not static; changes in parts of it must give rise to
the variable continuum polarization which is seen. The fact that we
have evidence for a variation in the net polarization over a $\approx
100$ day interval in the observed frame ($\approx 30$ day interval
in proper time) suggests that the size scales involved which lead
to changes in polarization are $< 10^{17}$ cm.  However, each
part of this asymmetric scattering region by itself would be expected
to give rise to highly-polarized continuum light ($p_{scatt,cont}
> 10$\%), but when averaged over the entire asymmetric region there
would be a much smaller net polarization.

For the purpose of illustration we note that if $\approx
80$\% of the flux of component D was composed of the continuum plus
BELs (as seen in the nearly identical spectra of components A, B
and C), and the remaining $\approx 20$\% of the flux of component D
was solely composed of microlensed scattered continuum (i.e. no BEL
flux), then the polarization of the scattered continuum component
would have to be $p_{scatt,cont} \approx 13$\% at $PA \approx 117$
deg to match the observations. This degree of polarization is
reasonable for scattering.

Owing to the fact that the BEL equivalent widths in component D
are observed to be smaller in comparison to the other components,
the observed microlensed scattered-light region evidently does not
scatter appreciable BEL flux. This suggests that the size-scale
of the polarized scattered-light region ($R_{scatt}$) must be less
than the size-scale of the region producing BELs, i.e., $R_{scatt} <
10^{18} L_{46}^{0.5}$ cm.  At the same time, in order to produce the
observed change in the polarization of component D relative to the
other three components, the polarized scattered-light region must
lie beyond the inner continuum-producing region which it reflects,
far enough so that the inner continuum-producing region is not
microlensed.  This is because, if the entire scattering region
were to lie within the microlensed region, there would be a near
constant magnification across the region during microlensing, the
continuum light and scattered continuum light would be similarly
amplified, and the polarization would remain unchanged among the
four components during microlensing.  Note that microlensing of an
unpolarized central continuum-producing region (e.g. the continuum
from the thermal accretion disk) is also not a possibility,
since this would reduce the polarization in component D, which is
not observed.  These constraints are new; they were not addressed
in the discussion of Chae \& Turnshek (1999) because component
polarization information was unavailable.

The size of the microlensed region on the source plane will
be of order the Einstein ring size. Following Chae \& Turnshek
(1999), the Einstein ring size on the source plane produced by
a microlensing star is given by $\eta_0 \approx 2 \times 10^{16}
(M/M_{\odot})^{0.5} h_{75}^{-0.5}$ cm, where $M$ is the mass of the
microlens. This is $\approx 8$ light-days for a solar mass star,
which is larger than the expected size of the continuum emitting
region from any accretion disk. We would expect the size of the
microlensed polarized scattering region to lie beyond this region.

Taken together our results therefore suggest that the size-scale
of the polarized scattered-light region in the Cloverleaf is
\begin{equation}
2 \times 10^{16} (M/M_{\odot})^{0.5} h_{75}^{-0.5} < R_{scatt} <
10^{18} L_{46}^{0.5} \ \ cm.
\end{equation}

From these results it is clear that a more rigorous future
program dedicated to monitoring (photometric, spectroscopic,
and polarimetric) the Cloverleaf would hold promise for providing
valuable constraints on models of the inner regions of QSOs (i.e
the BAL region, the BEL region, and the scattered-light region
producing the polarization).

\bigskip

We are grateful for helpful communications throughout the course of this 
work from Drs. S. Casertano, J. Krist, P. Massey, E. Monier, 
and J. Walsh. We also thank Dr. S. Mao for helpful discussions regarding 
microlensing and comments on the manuscript.

\newpage
\centerline{\bf Table 1.
WFPC2 Photometric and Polarimetric Observations}
\tabskip=0.5em
\halign to 
\hsize{\hfil#\hfil&\hfil#\hfil&\hfil#\hfil&\hfil#\hfil&\hfil#\hfil&\hfil#\hfil
\cr
\noalign{\vskip6pt\hrule\vskip3pt\hrule\vskip6pt}
 & Observation & Exposure   & Filter,   & PA\_V3 & Relative     \cr
 & Number      & Time (sec) & Polarizer & (deg)  & Orient (deg) \cr
\noalign{\vskip6pt\hrule\vskip6pt}
 &     &      & 15-16 Mar 1999 &      & \cr
\noalign{\vskip6pt\hrule\vskip6pt}
 &  1   & $8 \times 40$ & F555W, none &  57  & N/A  \cr
 &  2   & $8 \times 40$ & F555W, none &  98  & N/A  \cr
 &  3   & $8 \times 40$ & F555W, none & 100  & N/A  \cr
 &  4   & $1 \times 70$ & F702W, none &  57  & N/A  \cr
 &  5   & $2 \times 70$ & F702W, none &  98  & N/A  \cr
 &  6   & $1 \times 70$ & F702W, none & 100  & N/A  \cr
 &  7   & $8 \times 100$ & F555W, POLQN18  &  98  &  14 \cr
 &  8   & $8 \times 100$ & F555W, POLQN18  & 100  &  16 \cr
 &  9   & $8 \times 100$ & F555W, POLQ     &  98  &  77 \cr
 & 10   & $8 \times 100$ & F555W, POLQP15  &  98  &  92 \cr
 & 11    & $8 \times 100$ & F555W, POLQN33  &  57  & 138 \cr
 & 12   & $8 \times 100$ & F555W, POLQN33  &  98  & 179 \cr
\noalign{\vskip6pt\hrule\vskip6pt}
 &     &                &  23-24 Jun 1999 &      & \cr
\noalign{\vskip6pt\hrule\vskip6pt}
 &  13  & $8 \times 40$ & F555W, none & 279  & N/A  \cr
 &  14  & $8 \times 40$ & F555W, none & 320  & N/A  \cr
 &  15  & $8 \times 40$ & F555W, none & 322  & N/A  \cr
 &  16  & $1 \times 70$ & F702W, none & 279  & N/A  \cr
 &  17  & $1 \times 70$ & F702W, none & 320  & N/A  \cr
 &  18  & $1 \times 70$ & F702W, none & 322  & N/A  \cr
 &  19  & $8 \times 100$ & F555W, POLQN33  & 279  &   0 \cr
 &  20  & $8 \times 100$ & F555W, POLQN18  & 322  &  58 \cr
 &  21  & $8 \times 100$ & F555W, POLQ     & 320  & 119 \cr
\noalign{\vskip3pt\hrule}
}

\begin{deluxetable}{cccccc}
\tablenum{\bf 2a}
\tablewidth{0pc}
\tablecaption{\bf Relative Magnitudes of the Cloverleaf 
Components at Two Epochs\tablenotemark{a}}

\tablehead{
\colhead{Filter} &  
\colhead{Epoch }
 & \colhead{$m_{\mbox{\scriptsize A}}$} 
 & \colhead{$m_{\mbox{\scriptsize B}}$}
 & \colhead{$m_{\mbox{\scriptsize C}}$} 
 & \colhead{$m_{\mbox{\scriptsize D}}$} 
}

\startdata
F555W  &  15-16 Mar 1999 &  $0.000 \pm 0.003$  &  $0.193 \pm 0.003$ &  $0.300 \pm 0.004$ & $0.292 \pm 0.003$ \\
       &  23-24 Jun 1999 &  $-0.025 \pm 0.003$  &  $0.188 \pm 0.003$ &  $0.281 \pm 0.004$ & $0.368 \pm 0.003$ \\
\cline{1-6}
F702W  &  15-16 Mar 1999 &  $0.000 \pm 0.007$  &  $0.171 \pm 0.007$ & $0.320 \pm 0.008$ & $0.385 \pm 0.008$  \\
       &  23-24 Jun 1999 & $-0.004 \pm 0.008$  &  $0.163 \pm 0.008$ & $0.344 \pm 0.009$ & $0.446 \pm 0.009$  \\
\enddata

\tablenotetext{a}{\ The magnitude of component A is set to zero for the March epoch.}
\end{deluxetable}

\begin{deluxetable}{ccccc}
\tablenum{\bf 2b}
\tablewidth{0pc}
\tablecaption{\bf Relative Magnitude Changes of the 
Components Over 100 Days}

\tablehead{
\colhead{Filter}
 & \colhead{$\Delta m_{\mbox{\scriptsize A}}$}
 & \colhead{$\Delta m_{\mbox{\scriptsize B}}$}
 & \colhead{$\Delta m_{\mbox{\scriptsize C}}$ }
 & \colhead{$\Delta m_{\mbox{\scriptsize D}}$}
}

\startdata
F555W & $-0.025 \pm 0.004$  & $-0.005 \pm 0.004$ & $-0.019 \pm 0.006$ & $0.076 \pm 0.004$ \\
F702W & $-0.004 \pm 0.011$ &  $-0.008 \pm 0.011$ & $0.024 \pm 0.012$ & $0.061 \pm 0.012$ \\
\cline{1-5}
F555W+F702W\tablenotemark{a}  & $-0.023 \pm 0.004$ & $-0.005 \pm 0.004$ & $-0.010 \pm 0.005$ & $0.074 \pm 0.004$ \\

\enddata

\tablenotetext{a}{\ The $\Delta m$ listed is the variance-weighted mean of the F555W and 
F702W results.}

\end{deluxetable}

\newpage

\centerline{\bf Table 3. F555W Relative 
Polarimetric Results for the Cloverleaf Components$^a$}
\small
\tabskip=0.7em
\halign to 
\hsize{\hfil#\hfil&#\hfil&\hfil#\hfil&\hfil#\hfil&\hfil#\hfil&\hfil#\hfil
&\hfil#\hfil \cr
\noalign{\vskip6pt\hrule\vskip3pt\hrule\vskip6pt}
 &  Parameter   &   A   &   B   &   C   &   D   &  
   $\overline{\mbox{ABC}}^b$ \cr
\noalign{\vskip6pt\hrule\vskip6pt}
        &              & & 15-16 Mar 1999 & & & \cr
\noalign{\vskip6pt\hrule\vskip6pt}
       &  $q$ (\%)  &  $-1.35\pm 0.43$  & $-1.42\pm 0.46$ & $-1.42\pm 0.46$
& $-2.65\pm 0.47$  &  $-1.39\pm 0.26$   \cr
        &  $u$ (\%)  & $0.78\pm 0.49$ & $1.76 \pm 0.52$ & $1.10\pm 0.53$ &
$-1.13\pm 0.53$  &  $1.19\pm 0.30$  \cr
   & $p$ (\%) &  $1.56\pm 0.45$ & $2.27\pm 0.50$ & $1.79\pm 0.49$ 
   & $2.88\pm 0.48$ & $1.83\pm 0.28$ \cr
&$p_{\mbox{\scriptsize corr}}$ (\%) $^c$ & 1.49& 2.21 & 1.72 & 2.84 & 1.81 \cr
   & PA (deg) $^c$ & $75.0\pm 8.8$ & $64.5\pm 6.1$ & $71.0\pm 8.0$ & $101.5\pm 5.2$
   & $69.7\pm 4.4$ \cr
     & $\Delta q_{\mbox{\scriptsize D,$\overline{\mbox{ABC}}$}}$ (\%) &  ---  &  ---  &  --- &  $-1.26\pm 0.54$ &  
       $\equiv 0$  \cr
     & $\Delta u_{\mbox{\scriptsize D,$\overline{\mbox{ABC}}$}}$ (\%) &  ---  &  ---  &  --- &   $-2.32\pm 0.61$  & 
       $\equiv 0$  \cr
\noalign{\vskip6pt\hrule\vskip6pt}
        &              & & 23-24 Jun 1999 & & & \cr
\noalign{\vskip6pt\hrule\vskip6pt}
      &  $q$ (\%)  &  ---  & --- & ---
      & $-0.91\pm 0.61$  &  $-0.88\pm 0.34$   \cr
     &  $u$ (\%)  & --- & --- & --- &
       $-0.43\pm 0.62$  &  $0.11\pm 0.35$  \cr
   & $p$ (\%) & ---  & --- & --- & $1.00\pm 0.61$ & $0.89\pm 0.34$ \cr
 & $p_{\mbox{\scriptsize corr}}$ (\%) $^c$ & --- & --- & --- & 0.79 & 0.82 \cr
       & PA (deg) $^c$ & --- & --- & --- & $103\pm 18$  & $87\pm 11$ \cr 
     & $\Delta q_{\mbox{\scriptsize D,$\overline{\mbox{ABC}}$}}$ (\%) &  ---  &  ---  &  --- &  $-0.03 \pm 0.70$ &  
       $\equiv 0$  \cr
     & $\Delta u_{\mbox{\scriptsize D,$\overline{\mbox{ABC}}$}}$ (\%) &  ---  &  ---  &  --- &   $-0.54\pm 0.71$  & 
       $\equiv 0$  \cr
\noalign{\vskip3pt\hrule}
}

\normalsize
\noindent
$^a$ Quoted errors are relative statistical errors. \\
$^b$ The combined result of components A, B, and C. \\
$^c$ We use the formula of Wardle \& Kronberg (1974) to correct the bias 
toward higher polarization degree in low S/N ratio data: 
\[ p_{\mbox{\scriptsize corr}} = p \left[ 1-\left(\frac{\sigma_p}{p}\right)^2
\right]^{1/2}. \]
Our derived errors in $PA$ are in good agreement with their general 
prescription for estimating the $PA$ errors under such conditions: 
$\sigma_{PA} \approx 28.65\sigma_p/p$ deg.

\newpage

\begin{figure}
\centerline{\epsfig{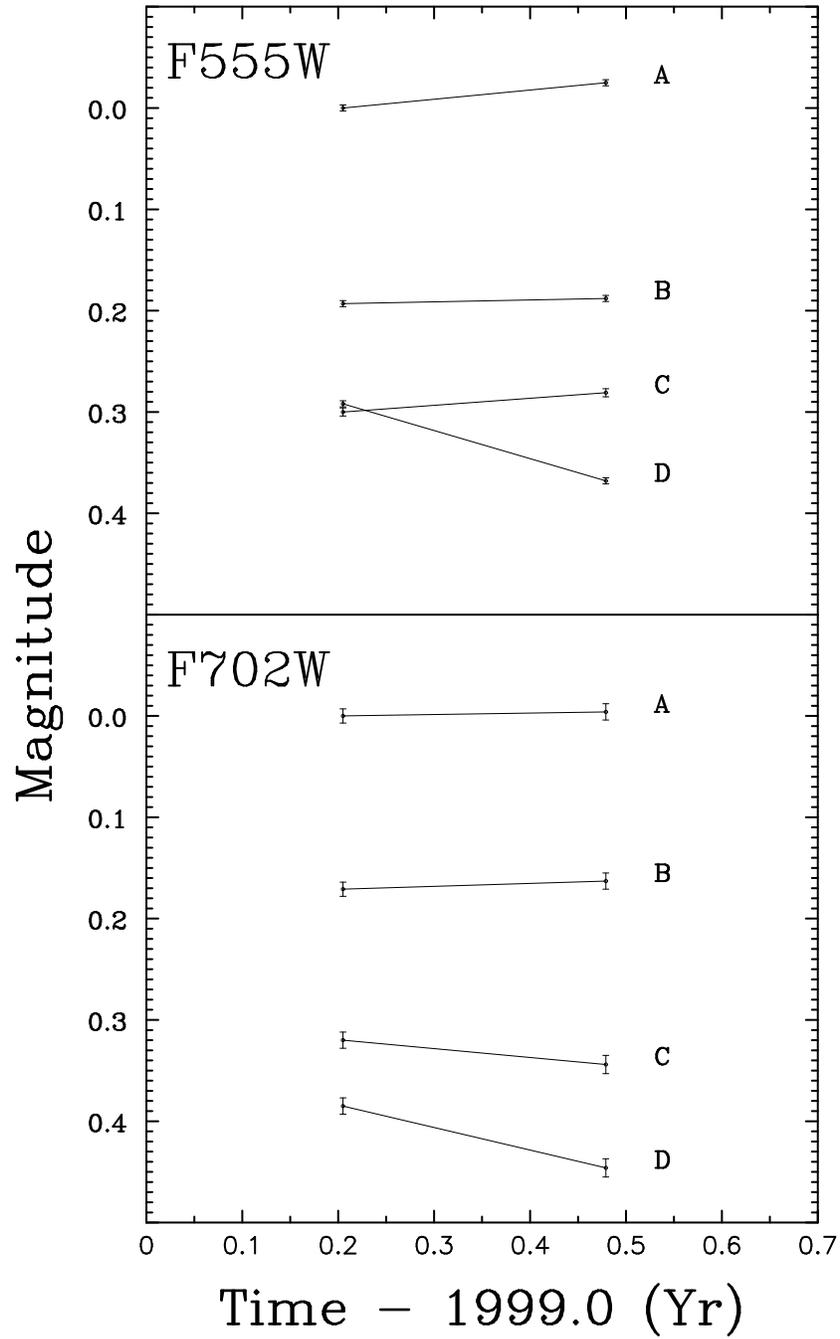}}
\caption{HST WFPC2 F555W and F702W brightnesses of Cloverleaf components
B, C, and D relative to component A in March 1999.}
\end{figure}

\newpage

\begin{figure}
\centerline{\epsfig{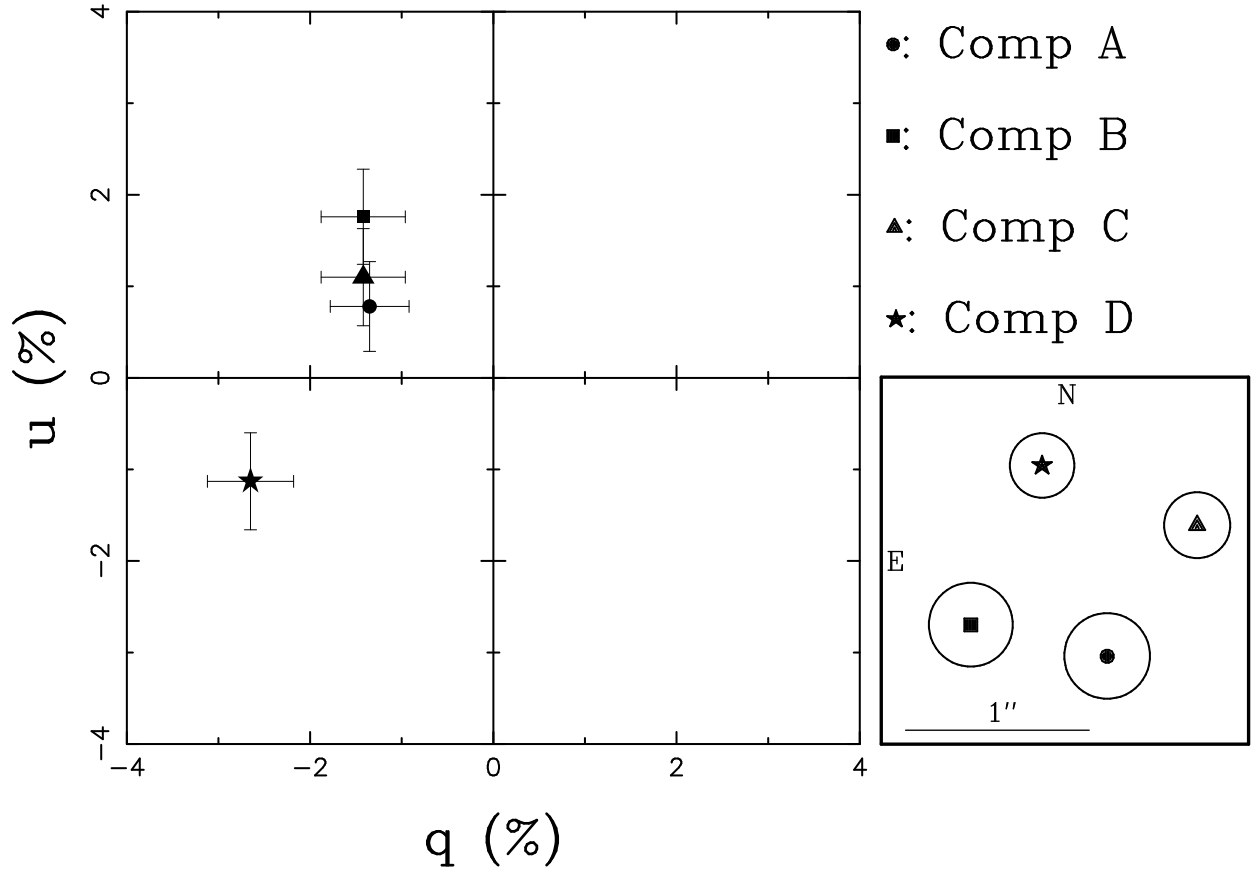}}
\caption{Stokes parameters derived from the March 1999 HST WFPC2
F555W observations of the four components of the Cloverleaf. The
legend on the upper right specifies the component
designations (A, B, C, D). The box on the lower right shows the 
relative spatial location of the components; the size 
of the circles drawn around the components is proportional to 
their relative brightnesses. The June 1999 observations
were less reliable (\S2), and for this reason they are not 
shown here but are only reported in Table 3.}
\end{figure}

\end{document}